\documentclass[12pt,preprint]{aastex}
\usepackage{epsfig}
\newcommand{\etal}{et~al.}

\newcommand{\kms}{\mbox{$\,$km s$^{-1}$}}

\newcommand{\dechms}[4]{$#1^{\rm h}#2^{\rm m}#3\mbox{$^{\rm s}\mskip-7.6mu.\,$}#4$}

\newcommand{\decdms}[4]{$#1^{\circ}#2'#3\mbox{$''\mskip-7.6mu.\,$}#4$}

\newcommand{\mdeg}[2]{$#1\mbox{$^\circ \mskip-7.6mu.\,$}#2$}

\newcommand{\HII}{\mbox{H\,{\sc ii}}}

\newcommand{\Htwo}{H$_{2}$}

\newcommand{\htco}{H$_{2}$CO}

\newcommand{\kmps}{km s$^{-1}$}

\shorttitle{}
\shortauthors{}

\received{}		

\begin{document}

\title{A comparison between anomalous 6-cm \htco\ absorption and CO(1-0) emission in the L1204/S140 region}

\author{M\'onica Ivette Rodr\'{\i}guez$^{1,2}$\\
Tommy Wiklind$^{1,3}$ \\
Ronald J. Allen$^{1}$ \\
Vladimir Escalante$^{2}$ \\
Laurent Loinard$^{2}$ \\[0.2in]}

\affil{$^{1}$Space Telescope Science Institute, 3700 San Martin Drive,
Baltimore, MD 21218, USA \\ monica, rjallen, wiklind@stsci.edu}

\affil{$^{2}$Centro de Radiostronom\'{\i}a y
Astrof\'{\i}sica, Universidad Nacional Aut\'onoma de M\'exico,
Apartado Postal 72 - 3, C.P. 58091, Morelia, Michoac\'an,
M\'exico\\ m.rodriguez, l.loinard, v.escalante@astrosmo.unam.mx
}

\affil{$^{3}$Affiliated with the Space Sciences Department of the European Space Agency}

\begin{abstract} We report observations of the dust cloud L1204 with
the Onsala 25-m telescope in the 6 cm (1$_{11}$-1$_{10}$) transition
of \htco. The observed region includes the
 S140 H${\alpha}$ arc. This spectral line is seen here in absorption
 against the cosmic microwave background, indicating the presence of
 widespread warm molecular gas at intermediate densities. Overall, the
 distributions of \htco\ and CO (taken from the literature) are fairly
 similar, though significant differences exist at small scales. Most
 notably, while the CO peak is nearly coincident with the S140
 H${\alpha}$ arc, the maximum \htco\ absorption is clearly separated
 from it by a full 10$'$ beam ($\sim$ 3 pc). We argue that these
 differences result from differing abundances and excitation
 requirements. The CO(1-0) line is more optically thick and more
 biased towards warm gas than the \htco\ 6 cm line. On the other hand,
 formaldehyde is more easily photodissociated and is, therefore, a
 poorer tracer of the molecular gas located immediately behind Photon
 Dominated Regions. \end{abstract}


\section{Introduction}
\label{sec:introduction}

Since \Htwo\ --the most abundant molecule in space-- lacks a permanent
dipole moment, its rotational transitions are prohibited. Although
the quadrupolar transitions exist, they are of little use for the
syudy of the bulk of molecular gas in the ISM because they require
high temperature to be excited. Instead, the structure and properties of
cold molecular clouds in the interstellar medium are usually studied
using low-energy rotational transitions of simple non-symmetric polar
molecules. For practical reasons, the first rotational transition (J =
1$\rightarrow$0) of carbon monoxide (CO), at 115.27 GHz has been the
most popular choice. This transition, however, has long been known to
be nearly always optically thick, so --for a given filling factor--
its intensity is expected to increase monotonically with the kinetic
temperature of the emitting gas. Clearly, this could have adverse
effects on efforts to establish the distribution of molecular gas in
the ISM from CO observations alone, because very low temperature gas
might go unnoticed in sensitivity-limited CO observations while warm
regions ($\gtrsim$ 20 K) will stand out even if they are not those
with the highest molecular content. The sources where these effects
might be most noticeable are those with large temperature gradients;
for instance in molecular clouds located in the immediate vicinity of
hot stars.

While all molecular emission tracers share this temperature dependance
to some degree, {\em absorption} lines can be detected even in very
cold gas, provided sufficiently bright background continuum sources
are available. The scarcity of such sources at the wavelength of the
common molecular tracers, however, has limited the usefulness of
absorption measurements in the study of specific Galactic molecular
clouds (e.g Evans et al. 1980). The 6-cm (1$_{10}$-1$_{11}$)
transition of ortho formaldehyde (\htco) offers an interesting
alternative. Owing to collisions with neutral particles that
selectively overpopulate the lower energy level, the excitation
temperature of the 1$_{10}$ $\rightarrow$ 1$_{11}$ transition lies
below 2.7 K \citep{toch69}. This allows the transition to be observed
in absorption against the cosmic microwave background (CMB)
\citep{syn69}, and makes it a potentially powerful tracer of molecular
gas in any direction of the sky.  The excitation requirements are such
the the 6 cm \htco\ line is a good indicator of the presence of cool
to warm molecular gas (T $\gtrsim$ 10 K) at intermediate densities
(10$^3$ cm$^{-3}$ $\leq$ n $\leq$ 10$^5$cm$^{-3}$). Unfortunately, the
absorption line is weak, so very large amounts of telescope time are
required to map large areas of the sky.

Recently, \citet{r06} conducted a blind search for \htco\ absorption
and compared CO emission and \htco\ absorption profiles towards the
Galactic anticenter. They found a rough, large-scale correlation
between these two tracers, and concluded that both lines
preferentially trace warm and dense molecular gas. Here, we will
examine this relation between CO and H$_2$CO at a somewhat smaller
scale using observations of the well-known, nearby star-forming region
Sharpless 140 (S140 --Sharpless 1959) associated with the dark dust
cloud Lynds 1204 (L1204 --Lynds 1962). L1204 is centered at $l$ =
\mdeg{107}{47}, $b$ = +\mdeg{4}{82} and covers an area of 2.5 square
degrees (Lynds 1962). At its southwest edge lies S140, a prominent
compact arc-shaped \HII\ region with an angular size of $\sim$ 2$'$
$\times$ 6$'$. The ionization of S140 is maintained by the nearby B0V
star HD211880 \citep{bl78}. The distance of S140/L1204 deduced from
the brightness of the exciting star is 910 pc \citep{c74}. S140 has
been the subject of many observational studies, that have usually
focused on the Photon Dominated Region (PDR) on the edge of L1204, and
on the embedded infrared sources located right behind it (e.g.,
Preibisch et al. 2001; Hayashi \& Murata 1992; Preibisch \& Smith
2002; Bally et al. 2002). Remarkably, while the dust cloud is seen as
an extended dark feature covering more than two square degrees, the CO
emission peaks immediately behind the H$\alpha$ arc (Heyer et
al. 1996; Evans et al. 1987; Blair et al. 1978), while only relatively
faint emission extends deep within the dust cloud \citep{hb97}. The
6-cm line of \htco\ was detected in absorption against the CMB in
L1204 near S140 by \citet{bl78} with the NRAO 43 m telescope, and
unexpectedly by \citet{e87} during VLA observations of the bright
condensation just northwest of S140. However neither of those studies
provided a extensive mapping of the \htco\ CMB absorption in L1204,
and the exact extension of the gas traced by \htco\ remains
unclear. In this article, we will present such a extensive mapping of
the 6 cm CMB absorption of \htco\ over most of the large dust complex
L1204, and compare our results with existing CO observations taken
from the literature.

\section{Data}
\label{sec:observations}

The \htco\ observations were obtained during two sessions (January and
September-October 2004, respectively) with the 25.6-m telescope of the
{\em Onsala Space Observatory} (OSO) in Sweden. At 6 cm, the angular
resolution is 10$'$, and our pointing precision was always better than
20$''$. Frequency-switching, with a frequency throw of 0.4 MHz was
used, and both polarizations of the incoming signal were recorded
simultaneously in two independent units of the autocorrelation
spectrometer. Each of these units provided 800 2 kHz-wide channels. At
the observed frequency of 4829.660 MHz, this setup provided a total
bandwidth of 99 km s$^{-1}$ and a (Hanning-smoothed) velocity
resolution of 8 kHz $\equiv$ 0.49 km s$^{-1}$. The spectrometer was
centered at the systemic velocity of S140, V$_{LSR}$ = $-$8.0 km
s$^{-1}$. Daily observations of the supernova remnant Cas A were used
to check the overall performance of the system. The system temperature
during our observations varied from 33 to 36 K.

In order to map the entire region behind S140, we observed 72
positions on a regular square grid with a 10$'$ spacing, centered at
$l$ = \mdeg{107}{0}, $b$ = +\mdeg{5}{3}; the resulting map uniformly
covers a \mdeg{1}{0} $\times$ \mdeg{1}{8} rectangular region (Fig.\ 1)
. The off-line data reduction was done with the CLASS program of the
GILDAS software package \citep{gf89}, and involved only the
subtraction of (flat) baselines from individual integrations and the
averaging of all spectra taken at the same pointing position. The
total integration time for each of these positions was about 10 hours,
yielding a typical final noise level of 3 mK (T$^*_A$).

The distribution of radio continuum sources in the region of L1204 has
been studied in detail by Allen Machalek \& Jia (in preparation), using data from the Canadian
Galactic Plane Survey. Fairly bright continuum emission is associated
with the H$\alpha$ arc and the embedded massive protostars located
behind it --but, as we will see momentarily no formaldehyde was
detected from either of these regions. In addition, a number of
extragalactic background sources as well as diffuse emission
associated with the dust cloud L1204 itself contribute to the overall
radio continuum. The typical brightness temperature average over the
Onsala beam at 6 cm, however, is only about 0.2 K, except towards the
H$\alpha$ arc and the embedded massive protostars (where again, no
absorption was detected). Since the brightness temperature is so
small, any \htco\ absorption profiles features detected must be
absorption of the cosmic microwave background radiation at 2.7 K.

In the analysis of our new observations, we will also make use of
$^{12}$CO(1-0) observations of L1204/S140 kindly provided by Dr.\
Tamara Helfer, and published in Heyer et al.\
(1996), and Helfer \& Blitz (1997). These data were obtained with the
14-m telescope of the {\em Five College Radio Astronomy Observatory}
(FCRAO) in Amherst (MA), and have an intrinsic angular resolution of
45$''$. For comparison with our formaldehyde data, we have
smoothed the CO(1-0) observations to 10$'$, and resampled them on our
observing grid.

\begin{deluxetable}{lcccr}
\tablewidth{0pt}
\tablecaption{Source positions.
\label{table:positions}}
\tablecolumns{5}
\tablehead{
\colhead{Source} &
\colhead{Position (\textit{l}, \textit{b})} &
\colhead{Size ($\Delta\alpha, \Delta\delta$)} &
\colhead{$\alpha$(J2000.0), $\delta$(J2000.0)} &
\colhead{Reference}
}
\startdata
L1204      & \mdeg{107}{37}, +\mdeg{4}{87} & \mdeg{1}{0} $\times$  \mdeg{2}{5} & \dechms{22}{26}{41} +\decdms{63}{15}{36}& Lynds (1962) \\
S140(H$_\alpha$) & \mdeg{106}{8}, +\mdeg{5}{3} & 2$'$ $\times$ 6$'$ &\dechms{22}{19}{23}+\decdms{63}{18}{16}& Sharpless (1959) \\
Our survey & \mdeg{107}{0}, +\mdeg{5}{3} & \mdeg{1}{0} $\times$ \mdeg{1}{8} & \dechms{22}{20}{52} +\decdms{63}{24}{49}& This paper \\
\enddata
\end{deluxetable}

\begin{figure*}[ht!]  
\epsscale{0.5}
 \plotone{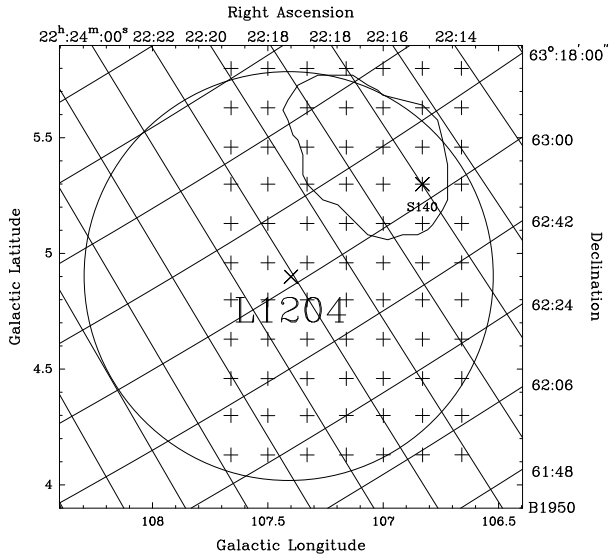} 
 \caption{This figure shows a sketch of the L1204/S140 region, with our observed
positions shown as "+" symbols. The correspondence between Galactic
coordinates (used throughout the paper) and equatorial coordinates
(that have usually been preferred for observations of S140) is
indicated. The circle represents the \mdeg{2}{5} size of the dust
cloud L1204 as reported by Lynds (1962). The central positions of
L1204 and S140 are shown as "\*" symbols. The contour represents 
the lowest value of the \htco\ absorption.   
\label{fig:sketch}}
\end{figure*}

\section{Results}
\label{sec:results}

\begin{figure*}[ht!]
\epsscale{0.6}
\plotone{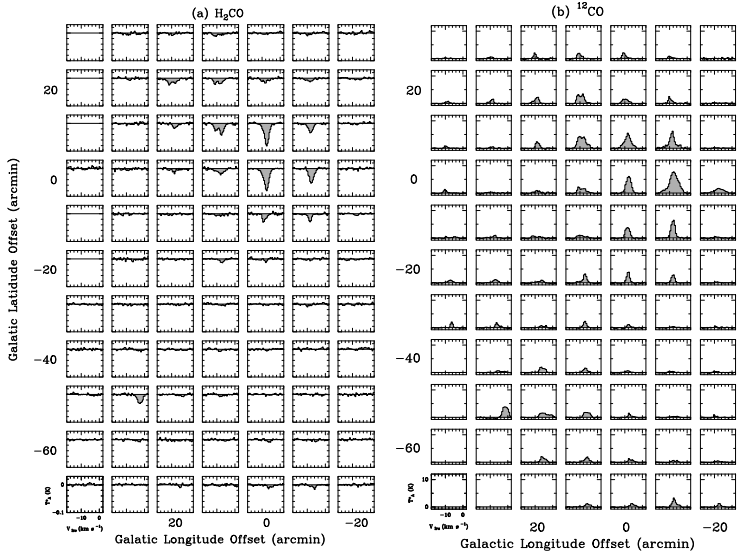}
\caption{(a) Mosaic of \htco\  CMB absorption spectra observed in the 
L1204/S140
region. The (0,0) position corresponds to $l$ = \mdeg{107}{0}, $b$ =
\mdeg{5}{3}, about 12$'$ east of S140, and about 40$'$
north-west of the nominal center of L1204 (Lynds 1962). (b) 
Corresponding CO(1-0) observations smoothed to 10$'$ (see
text). Note that the CO emission is located a full 10$'$ west of 
the maximum H$_2$CO absorption. 
\label{fig:mosaic}}
\end{figure*}

Formaldehyde absorption was detected in at least 16 of our 72 observed
positions (see Fig.\ \ref{fig:mosaic}). The maximum absorption is
located 10$'$ arcmin behind the S140 \HII\ region at a LSR velocity of $-$8.0 km
s$^{-1}$, similar to that of the CO emission detected in that area.  A
second spatio-kinematical structure is detected towards the
north-east (here, and in the rest of the paper, north and all other directions refer to
Galactic coordinates), at $V_{LSR}$ $\sim$ $-$11 km s$^{-1}$. Both components are
presumably associated with L1204, and have clear CO counterparts
(Fig.\ \ref{fig:mosaic} -- Blair et al. 1978; Evans et al. 1987;
Sugitani \& Fukui 1987; Park \& Minh 1995). There is also an isolated
absorption feature towards the southeast, at $V_{LSR}$ $\sim$ $-$2.5
km s$^{-1}$. Given its low LSR velocity, this feature is likely
unrelated to L1204, and is probably a local cloud along the line of
sight. Thus, while \citet{sf87} identified three molecular components
associated with L1204 in their $^{13}$CO observations, we only find
two in our formaldehyde data. We do find evidence, however, for a
systematic velocity gradient across the cloud. \citet{pm95} argued
that this complex overall spatio-kinematical morphology was created
when S140 and L1204 were swept up by an expanding shell associated
with the Cepheus bubble. Our data do not illuminate this assertion any
further, and a more thorough study is necessary to understand the
detailed structure of this region.

\begin{figure*}[ht!]  \epsscale{0.6} \plotone{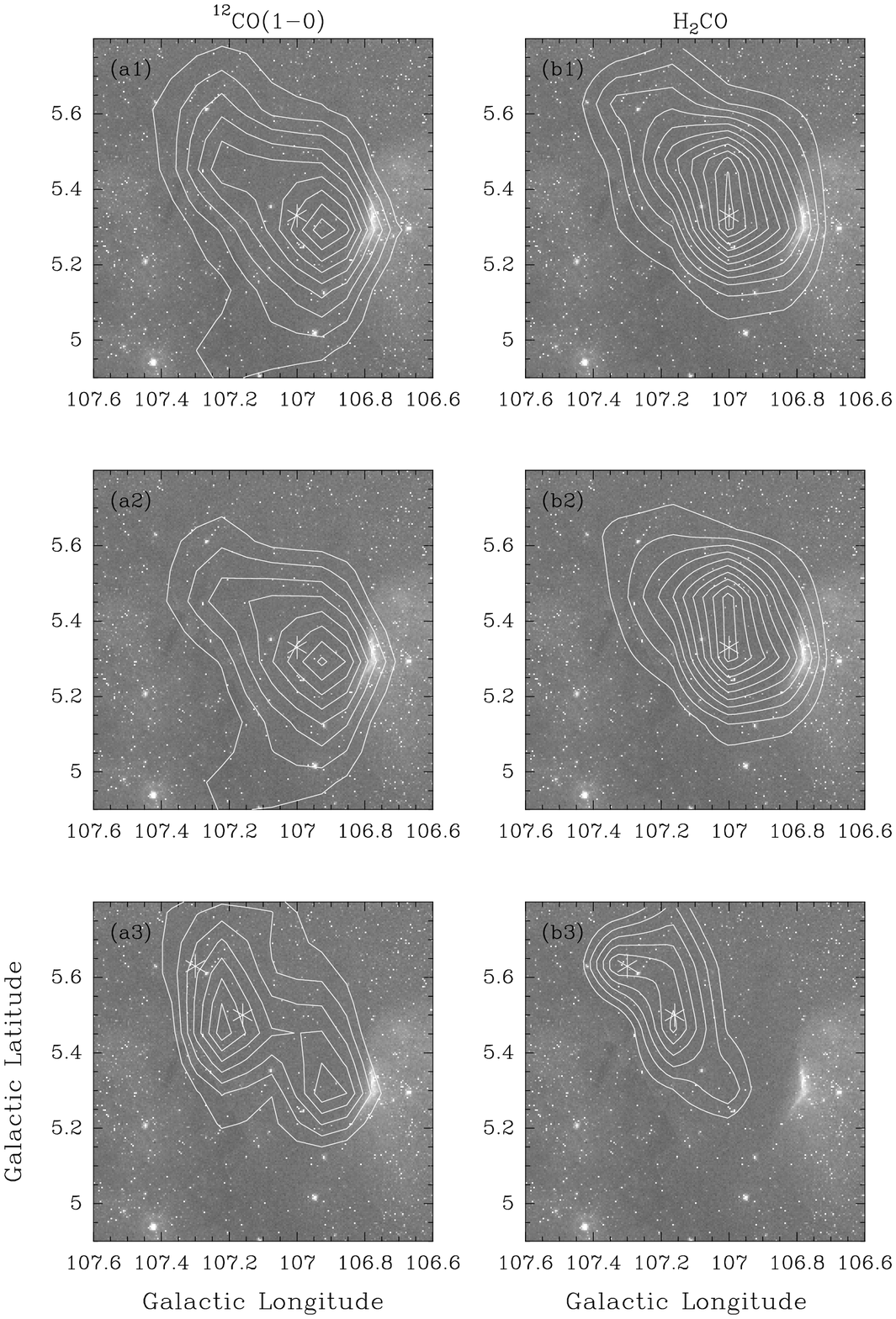} \caption{All
panels show a grey-scale version of the DSS-red image of the region
around \textit{l} = \mdeg{107}{0}, \textit{b} = \mdeg{5}{4}. The
H$_{\alpha}$ arc of S140 is clearly visible on these (red) images near
\textit{l} = \mdeg{106}{8}, \textit{b} = \mdeg{5}{3}. In the three
left panels (a1--a3), CO contours taken from the smoothed CO data of
\citet{he96} are overlaid on top of the DSS image, whereas in the
three right panels (b1--b3), our \htco\ contours are overlaid. The
contours in the top two panels (a1--b1) include the entire velocity
range associated with L1204 (from -12 to -5 K km s$^{-1}$), whereas in
the middle two (a2--b2) and bottom two (a3--b3) panels, the contours
correspond only to the -11 K km s$^{-1}$ and the -8 K km s$^{-1}$
components, respectively. The asterisks correspond to the position of
the peak \htco\ absorption for each component.  \label{fig:2comp}}
\end{figure*}

\section{Comparison with other molecular tracers}
\label{sec:comparison}

The S140/L1204 region has been observed in many different molecular
tracers (e.g.\ Tafalla et al.\ 1993, Zhou et al. 1993, Park \& Minh
1995), but most of these observations have focused either on the S140
PDR or on the embedded infrared sources located just behind S140,
while only a few observations covered the entire dust cloud. Indeed,
the first CO observations of S140 \citep{bl78} only covered a limited part  
of the region. To our knowledge, the only existing
large-scale CO map of L1204 is that obtained in the 90s with the FCRAO
telescope (see \S 2) and published by Heyer et al.\ (1996) and Helfer
\& Blitz (1997)\footnote{The region lies on the edge of, and is only
partly covered by, the CfA Galactic plane survey of Dame et al.\
(2001).}. As mentioned earlier, we will use a smoothed version of that
dataset here in order to compare with our formaldehyde observations.

In general, the CO emission and \htco\ absorption morphologies in this
region are quite similar (Fig.\ \ref{fig:2comp}). This was already
noticed by \citet{bl78} in their 6$'$ observations. It is also in good
agreement with the results obtained towards the Galactic anticenter by
Rodr\'{\i}guez et al.\ (2006), and towards the Orion molecular complex
by Cohen et al. (1983). There are, however, several noteworthy
differences between the CO emission and \htco\ absorption in S140. The
first difference is the fact that the CO peak and the \htco\
absorption maximum are not located at the same position. The CO
integrated intensity map (Fig.\ \ref{fig:2comp}, see also Fig.\
\ref{fig:mosaic}) shows that the maximum CO emission occurs just
behind the S140 H$\alpha$ arc at the western edge of L1204, while only
comparatively fainter emission extends to greater longitudes. The
maximum \htco\ absorption, however, is located about 10$'$ eastward of
the CO peak. A second notable difference between the CO and \htco\
profile is the existence of a CO emission "tail" in the
south/southeast part of the main cloud with little or no \htco\
counterpart (Figs. 2 and 3).

\begin{figure*}[ht!]  \plotone{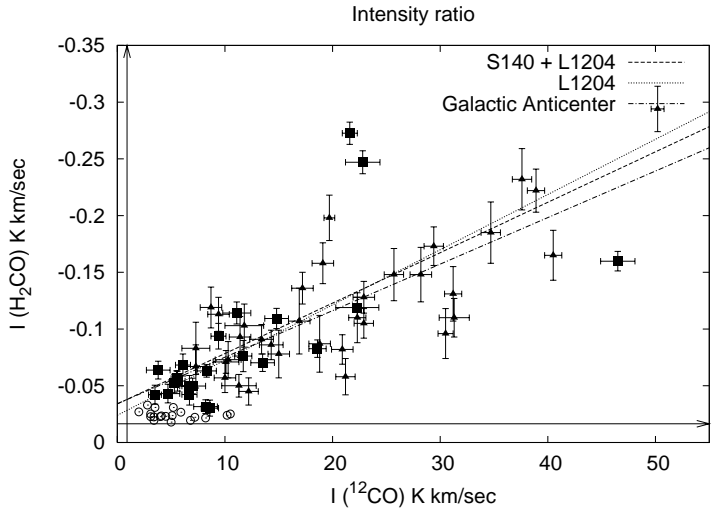} \epsscale{1.0}
\caption{Correlation between the \htco\ absorption line intensity and
the $^{12}$CO(1-0) emission line intensity at corresponding points in
L1204. The squares correspond to data from Table \ref{table:mom}, the
open circles correspond to \htco\ upper limits and the triangles
correspond to the Galactic anticenter data previously published in
\citet{r06}. The horizontal and vertical lines show the ``best case''
detection limit. The dashed line is the least-squares fit for the
L1204/S140 region data, and the dash-dotted line is the least-squares
fit for the Galactic anticenter data from Rodr\'{\i}guez et
al. (2006). Note that the fits do not pass through the (0,0) point, suggesting that the 
relation is not linear at low intensity values.
\label{fig:intensityratio}} \end{figure*}

In order to study the relation between \htco\ absorption and CO(1-0)
emission in a more quantitative way, we have computed the moments of
the profiles shown in Fig.\ \ref{fig:mosaic}. The results are listed
in Table \ \ref{table:mom} (Appendix B). When two velocity components
are visible at a given pointing, the moments for each were computed
separately. Intensities above $3\sigma$ are shown as squares in Fig.\
\ref{fig:intensityratio}, and were used to make least-square fits (see
below). The two spatio-kinematical components that we identified in
our formaldehyde dataset behave quite similarly with respect to the
CO-\htco\ relation, and are plotted together in Fig.\
\ref{fig:intensityratio}. The best least-squares fit to a straight line for 
the entire L1204 dataset yields:

\begin{equation}
I(H_2CO) = (4.4 \pm 1.1) \times 10^{-3} ~~ I(CO) + (34 \pm 16) \times 10^{-3} \mbox{~K km s$^{-1}$}.\label{eq:relation1}
\end{equation}

\noindent
We shall see momentarily that the CO emission near S140 may be
particularly bright because of local heating. Ignoring the pointings
very near S140, however, yields a fairly similar relation between CO
and \htco:

\begin{equation}
I(H_2CO) = (3.8 \pm 0.9 ) \times 10^{-3} ~~ I(CO) + (31 \pm 9) \times 10^{-3} \mbox{~K km s$^{-1}$}.\label{eq:relation2}
\end{equation}

\noindent
The small difference between these relations presumably reflects the
differing excitation requirements for the two lines. The relation
between CO and \htco\ given by Eqs.\ \ref{eq:relation1} and
\ref{eq:relation2} for the L1204 region is almost identical to that
found towards the Galactic anticenter by Rodr\'{\i}guez et al.\
(2006):

\begin{equation}
I(H_2CO) = (4.1 \pm 0.5 ) \times 10^{-3} ~~ I(CO) + (34 \pm 13) \times 10^{-3} \mbox{~K km s$^{-1}$}.\label{eq:relation3}
\end{equation}

It is important to note, however, that, in spite of the agreement
between the fits to the Galactic anticenter and S140 data, there is
very significant scatter in the CO-\htco\ relation, some points lying
nearly 10$\sigma$ away from the linear relation. This situation was
already noticed by \citet{r06} in their study of the Galactic
anticenter. This lack of a detailed correspondence between CO emission
and \htco\ absorption presumably reflects differences in the
excitation conditions of the two tracers, as we will now discuss in
the next section.

\section{Discussion}
\label{sec:discussion}

The comparison between \htco\ absorption and CO(1-0) emission profiles in 
the Galactic Anticenter and in the L1204/S140 region has led us to three important 
observational conclusions:

\begin{enumerate}

\item Qualitatively, the morphology of CO and \htco\ are quite similar, and
quantitatively, the line integrated intensities correlate quite well
with one another.

\item The scatter in the CO-\htco\ relation is, however, significantly 
larger than the observational errors.

\item In the specific case of S140, the CO emission peak is offset by
about 3 pc from the locus of the deepest formaldehyde absorption, and
there is a region south of the main cloud where significant CO
emission is detected with little or no \htco\ counterpart.

\end{enumerate}

From the general large-scale correspondance between the CO(1-0) and
\htco\ 6 cm integrated maps (Fig.\ \ref{fig:2comp}), and from the fair
correlation between their line intensities (Fig.\
\ref{fig:intensityratio}), we conclude that the physical conditions
needed for the excitation of both lines are quite similar. The
calculations presented in the Appendix A, indeed show that both lines
preferentially trace warm gas at intermediate densities (10$^{3.6}$$<$
n $<$ 10$^{5}$ for \htco; n $>$ 10$^{2.8}$ for CO). In this scheme,
the offset between the CO(1-0) peak and the \htco\ maximum absorption
may seem puzzling. Note that a similar trend is seen at higher
resolution: while the CO peak in the full-resolution CO map published
by Heyer et al.\ (1996) and Helfer \& Blitz (1997) is at $l$ = 106$^{o}$.8, $b$
= +5$^{o}$.3, the \htco\ absorption feature seen in the high-resolution VLA
images published by Evans et al.\ (1987) is centered around $l$ = 106$^{o}$.9, $b$
= +5$^{o}$.3, again a few arcminutes to the east. 

We suggest that a combination of two effects may explain this puzzling
result. First, it can be seen from the excitation analysis presented
in Appendix A that the \htco\ absorption strength "saturates" for $T
\gtrsim 30$ K, whereas the temperature of the CO emission continues to
rise at higher kinetic temperature (Fig.\ 5). For example, while the
\htco\ line strength increases by only about 30\% when the kinetic
temperature goes from 20 to 40 K, the CO(1-0) line intensity increases
by more than a factor of two. According to \citet{pm95}, the CO
brightness temperature is about 40 K at the peak and 20 K for the rest
of the cloud. Thus, the strong CO peak behind S140 may well be largely
due to enhanced kinetic temperatures related to local heating (by the
external star providing the ionization of S140, and/or by the infrared
sources embedded in the cloud). As one progresses into the cloud, the
local heating diminishes, and the CO line intensity fades. Also, it
should be pointed out that the formaldehyde calculations presented in
the Appendix show that the 6-cm line should be seen in emission rather
than absorption when the density exceeds 10$^5$ cm$^{-3}$. The fact
that this is not the case near the CO peak (neither in our low-resolution data, 
nor in the high-resolution VLA data presented by Evans 1978) suggests that 
the gas density there is lower than 10$^5$ cm$^{-3}$.
Other effects that could explain the offset between the CO and the H$_2$CO 
peaks are the lower dissociation energy, and the lower abundance 
(and, therefore, lower self-shielding) of formaldehyde
compared to CO (see Appendix A.3). In a photo-dissociated region, these 
effects should combine to create a stratified distribution where CO survives nearer 
the source of the UV photons than H$_2$CO.
This stratification, combined with the heating of the CO, would
naturally lead to the offset between H$_2$CO and CO seen in the
present data.

Finally, the origin of the other main difference between CO and \htco\
in S140, namely the existence of CO emission at the south of L1204
with no or little formaldehyde counterpart, is likely related to
another aspect of the excitation differences between the 6 cm line of
formaldehyde and the 1-0 transition of carbon monoxide. Fig.\ 6 of the
Appendix A.2 shows that the density detection limit for \htco\ line is
$\sim$10 times larger than the density limit for the CO(1-0) line. We
therefore suggest that the gas traced by the CO emission to the south
of L1204 is of relatively very low density. It is interesting to note,
indeed, that classical high-density molecular tracers (e.g.\ CS or
NH$_3$) have only been detected around the CO peak behind S140, and
not in the southern region of the cloud.

Thus, we conclude that the CO(1-0) and \htco\ 6 cm lines both tend 
to preferentially trace warm gas at intermediate densities. There are, 
however, significant differences related either to differing excitation 
requirements or to differing abundances. These differences can easily 
explain the large scatter in the CO--H$_2$CO relation.

\section{Conclusions}
\label{sec:conclusions}

The main conclusions of this work are the following:

\begin{enumerate}

\item We have mapped a large region (70$'$ $\times$ 110$'$) around
L1204/S140 in the 6 cm line of formaldehyde, observing a total of 72
regularly-spaced positions every 10$'$ on a regular grid. The center
of our map was at $l$ = \mdeg{107}{0}, $b$ = +\mdeg{5}{3}, and
formaldehyde was detected against the cosmic microwave background in
at least 16 of our 72 positions (Fig.\ 2).

\item
The formaldehyde emission can be separated in three spatio-kinematical
components (Fig.\ 3): two (at $V_{LSR}$ $\sim$ --11 km s$^{-1}$ in the northeast 
part of the cloud, and at $V_{LSR}$ $\sim$ --8 km s$^{-1}$ just behind S140) are 
clearly associated with L1204, whereas the other (an isolated component at 
$V_{LSR}$ $\sim$ --2.5 km s$^{-1}$ towards the southeast) is most likely a local 
foreground cloud unrelated to S140/L1204.

\item
Both qualitatively and quantitatively, the CO(1-0) emission and the
formaldehyde 6 cm absorption lines correlate fairly well. An
excitation analysis shows that both preferentially trace warm gas at
intermediate densities.

\item
There are, however, notable differences between the CO and \htco\
lines, that can be traced to differing excitation requirements and
abundances. Those differences are most likely the origin of the 
large scatter in the CO-\htco\ intensity correlation. 

\end{enumerate}

\acknowledgements

We thank Professor Roy Booth, director (retired) of the radio
observatory at Onsala, for generous allocations of telescope time and
for his warm hospitality during our several visits to the
observatory. We are also grateful to the observatory technical and
administrative staff for their capable assistance with our observing
program. We thank Tamara Helfer for supplying us with the CO data cube of S140. We
acknowledge the financial support of the \textit{Direcci\'on General de
Asuntos del Personal Acad\'emico} (DGAPA), \textit{Universidad Nacional
Aut\'onoma de M\'exico} (UNAM) and \textit{Consejo Nacional de Ciencia y
Tecnolog\'{\i}a} (CONACyT), in M\'exico, and the Director's
Discretionary Research Fund at the Space Telescope Science
Institute. The Digitized Sky Surveys were produced at the Space
Telescope Science Institute under U.S. Government grant NAG
W-2166. The images of these surveys are based on photographic data
obtained using the Oschin Schmidt Telescope on Palomar Mountain and
the UK Schmidt Telescope. The plates were processed into the present
compressed digital form with the permission of these institutions.

\appendix

\section{Model calculations}
\label{sec:model}

\subsection{Collisional pumping}
\label{subsec:collisional}

Observations of \htco\ in dark clouds show that the anomalous
absorption of the 6 and 2~cm lines is due to collisions with \Htwo\
that selectively overpopulate the lower levels of the lines
\citep{e75}. We calculated the non--LTE equilibrium populations of
the first 40 levels of ortho--\htco\ assuming excitation by the 2.7~K
background and collisions with \Htwo. \citet{gr91} has calculated
excitation rates of these levels for collisions with He taking
advantage of the spherical symmetry of the He potential for kinetic
temperatures $T=10$ to 300~K. According to \citet{gr91}, excitation
rates by \Htwo\ collisions could be 2.2 times higher than those by He
because of the smaller reduced mass and differences in the interaction
potentials.  We will show the results of the calculations under the
assumption that the \Htwo--\htco\ collisional rates are the same as
the He--\htco\ rates. Probabilities for the radiative transitions were
taken from \citet{jarus86}.

The optical depths of the transitions involved in the pumping mechanism 
are generally larger than 1 at high densities and a radiative transfer 
calculation is required.  Two limiting approximations in the radiation transport are often
considered in molecular clouds: the large velocity gradient (LVG)
model and the microturbulent model \citep{leli76}.  The LVG model
assumes that the line profile is dominated by systematic motion of the
gas while the microturbulent model assumes that the turbulent velocity
is much larger than any systematic motion.  S 140 is likely to have
several velocity components, but the existence of large systematic
motions in molecular clouds and the validity of the LVG model has not
been well established in other molecular clouds (e.g. \citet{e75},
\citet{zue97}, \citet{z90}).  We therefore use the microturbulent model, and for simplicity we will 
use the escape probability formalism to account for photon trapping in a turbulent medium in 
plane-parallel slab of mean total optical depth $\tau_t$ that is
perpendicular to the line of sight.  In this case the A--value of a
transition in the equations of statistical equilibrium is multiplied
by a ``loss probability'' $P(\tau,\tau_t)$ that depends on the mean
optical depth $\tau$ in the slab.  There are many different ways to
define $P(\tau,\tau_t)$, which can differ by several orders of
magnitude for large optical depths.  We will use the form suggested by
\citet{hust92} for a uniform medium with no continuum absorption:
\begin{equation}
P(\tau,\tau_t)=\onehalf[K_2(\tau)+K_2(\tau_t-\tau)]\ ,
\label{escape}
\end{equation}
where
\begin{equation}
K_2(\tau)=\int_{-\infty}^\infty dx\,\phi(x)E_2[\tau\phi(x)]\ ,
\end{equation}
and $E_2$ is the second exponential integral function.
The function $K_2(\tau)$ can be calculated from
fits by \citet{hu81} for a normalized Doppler profile,
$\phi(x)=\exp(-x^2)/\sqrt{\pi}$.

Equation~(\ref{escape}) can be viewed as the single flight escape
probability through either side of the slab averaged over the line
profile.  The probability that a photon of an isotropic background
reaches optical depth $\tau$ in this slab is also given by
equation~(\ref{escape}), and the blackbody continuum is thus
attenuated by a factor $P(\tau,\tau_t)$.  The mean optical thickness
is given by
\begin{equation}
d\tau={\sigma_{lu}\over\Delta\nu_{D}}n_l\left(1-{n_u/g_u\over n_l/g_l}\right)dz\ ,
\label{tau}
\end{equation}
where $\sigma_{lu}$ is the absorption cross section of the transition
and $\Delta\nu_{D}$ is the Doppler width.  We assumed a Doppler width 
of 2\kms\ (FWHM = $2 \sqrt{ \ln 2} \Delta\nu_{D}$ = 3.3\kms). The level population 
and its statistical weight are given by $n$ and $g$ respectively with subindex $u$ for 
the upper and $l$ for the lower level. 

The emergent line brightness temperature with subtracted
background $T_0$ is found by direct integration of the
source function throughout the slab as
\begin{eqnarray}
\Delta T_b &=& T_b-T_0 \nonumber \\
&=&T_0\big[\exp(-{\cal T})-1\big] \nonumber \\ &+&{h\nu\over k}
\int_0^{{\cal T}}\exp(-{\cal T}+\tau)
\left({n_l/g_l\over n_u/g_u}-1\right)^{-1}\,d\tau \ ,
\label{tb}
\end{eqnarray}

where 

\begin{equation}
\tau_t=\int_0^L {\sigma_{lu}\over\Delta\nu_D}n_l
\left(1-{n_u/g_u\over n_l/g_l}\right)\,dz \ ,\label{totaltau} 
\end{equation}
is the total mean optical depth of a slab of thickness $L$ 
and ${\cal T}=\tau_t\phi(0)$  is the line--center total optical 
depth. 

\begin{figure*}[ht!]
\plotone{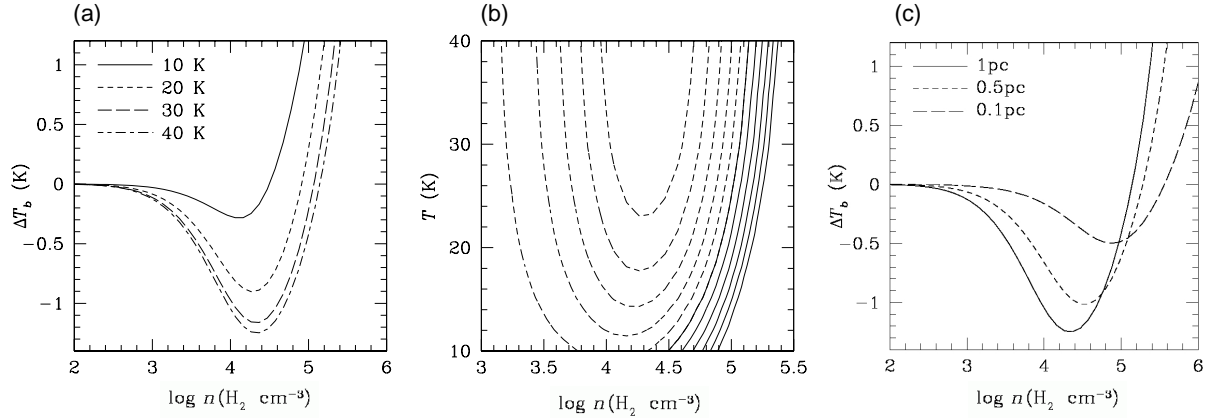}
\epsscale{0.8}
\caption{(a) Brightness temperature minus background continuum vs. density for
different kinetic temperatures $T$ of a 1--pc thick slab. (b) Contour intervals are 
0.2~K from $\Delta T_b=-1$ to 1~K. (c) Brightness temperature minus background 
continuum vs. density for different slab thicknesses at 
T = 40~K.
\label{fig:tbvsn}}
\end{figure*}

Equation~\ref{tb} gives the correct asymptotic limits: $T_b\to T_0$ when the 
density goes to 0 and $T_b\to T$ for high densities. The population densities 
and the optical depths as a function of position in the slab in 
equations~(\ref{escape}) and~(\ref{tau}) were calculated iteratively. A 
convergence of $10^{-3}$~K in $T_b$ was achieved after a few
iterations for $T\leq 40$~K and \Htwo\ densities $n({\rm
H_2})<10^6\,{\rm cm^{-3}}$. For higher densities and temperatures the
procedure becomes unstable. Fig.\ \ref{fig:tbvsn} shows the calculated
$\Delta T_b$ for a 1~pc--thick slab as a function of the
\Htwo\ density and a constant \htco\ abundance of $2\times10^{-9}$
with respect to \Htwo\ (\citet{ha92}, \citet{le84}). The assumed
thickness of the slab has an important effect in the anomalous
absorption as shown in Fig.\ \ref{fig:tbvsn}.  As the thickness of the
slab decreases, the effectiveness of the pumping mechanism that cools
the line decreases.

\citet{g75} identified some transitions,  
like $1_{11}\to 3_{12}$, that produce selection effects in the
excitation of some levels that cool the \htco\ doublets.  We have
tested our model for possible variations of the collision rates.  An
overall increase of collisional rates by a factor of 2.2 decreases
$\Delta T_b$ in figure~\ref{fig:tbvsn} by 0.3 to 0.5~K for $T\geq20$~K
and $n({\rm H_2})>1.6\times10^4\,{\rm cm^{-3}}$.  For lower $T$ and
$n({\rm H_2})$ there is very little variation in the predicted $T_b$.

\subsection{A PDR model for CO}

The UV field has a higher influence on the CO brightness temperature
than on the \htco\ brightness temperature. Both molecules are quickly
photodissociated near the edge of clouds, but the larger abundance of
the CO molecule makes its chemistry and interaction with radiation
more complex. In order to take into account the variation of CO
abundance along the line of sight, we used the Meudon PDR code to
calculate the CO brightness temperature of a plane--parallel slab
irradiated by a UV field \citep{leb93}. A detailed description of a
revised version of the code is given by \citet{lep06}.

The sharp separation between the molecular, atomic and ionized
emissions suggests that the L1204/S140 interface is a PDR viewed
nearly edge--on \citep{hm92} irradiated by HD~211880. The angle of
incidence of the star's radiation on the PDR boundary is an unknown
parameter but appears to be more-or-less perpendicular. Furthermore, 
the infrared embedded sources are probably young stars that may also 
enhance the radiation field \citep{e89}.

\begin{figure*}[ht!]
\plotone{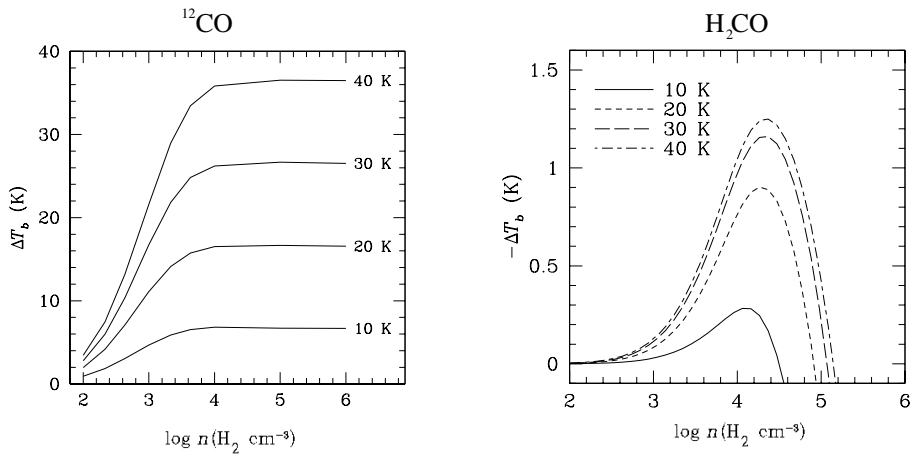}
\caption{Brightness temperature minus background continuum vs. density for
different kinetic temperatures $T$ for both the $^{12}$CO(1-0) line and the 6 cm 
\htco\ line (Note that $\Delta$T$_b$(k) is plotted "negative" compared to Figure \ref{fig:tbvsn}).
\label{fig:comparison}}
\end{figure*}

We ran the code with its parameters set to represent a plane--parallel
slab irradiated from one side by a UV field with an enhancement factor
$\chi=200$ with respect to the \citet{dr78} average interstellar
radiation field. In a PDR, the gas is heated by the
photoelectric emission from grains and PAH's, \Htwo\ formation in
grains, UV pumping in the Lyman and Werner bands, gas--grain
collisions, photoionization, and photodissociation. As the UV radiation
is absorbed deeper into the cloud, other processes like cosmic rays
and chemical reaction energies become important. Cooling is produced
by fine--structure and molecular line emission. Shocks and turbulence
can keep a PDR away from isobaric equilibrium. However in our model
the temperature and density of the PDR were kept constant in order to
compare the results with our \htco\ model in Fig.\
\ref{fig:comparison}. We used the chemical network given for S140 by
the Meudon group at its Internet
site\footnote{http://aristote.obspm.fr/MIS/pdr/exe.html}, which does
not include \htco. Thus the CO and \htco\ calculations represent
different models, and Fig.~\ref{fig:comparison} is given only as
indication of the local conditions that produce the emission and
absorption for each molecule.

\subsection{Photodissociation of \htco}
\label{subsec:photo}

The \htco\ molecule is quickly photodissociated into CO and \Htwo\ or
H in the average UV interstellar radiation field with a rate of $1.0
\times10^{-9}\exp(-1.7A_V)\,{\rm s^{-1}}$ \citep{vand88}. 
\citet{ke85} estimated that far--ultraviolet radiation (FUV)
from the star HD211880 will have an enhancement factor of
$G_0=150$ with respect to the average interstellar field
\citep{h68} at the ionization front, although \citet{sp97} 
found that a more intense radiation field may be needed to 
explain the \Htwo\ rotational emission. 
\htco\ has a dissociation energy of 3.61 $\pm$ 0.03 eV \citep{su86}, and 
a photodissociation rate of 1.0 $\times$ 10$^{-9}$ sec$^{-1}$ in the
interstellar field \citep{vand88} while for CO the values are 11.2 eV
and 2.0 $\times$ 10$^{-10}$ sec$^{-1}$ respectively.  Thus it is possible 
that the \htco\ will be selectively photodissociated near the S140
ionization front and the bright PDR region, where the CO emission
peaks. Detailed PDR model calculations by \citet{li02} show that at
20$'$ from the ionization front, where we observe the \htco\ maximum,
$G_0\leq20$ and $A_V\sim15$.

We added 62 reactions involving $\rm H_2CO$ and $\rm H_2CO^+$ 
taken from the UMIST data base (Woodall et al., 2007) 
\footnote{http://www.udfa.net/} to the chemical network of 
the Meudon group mentioned above and ran a PDR model with 
$G_0=200$, $T=40$K and constant density of $\rm 10^3\,cm^{-3}$. 
We found that $\rm H_2CO$ has significant abundance only at depths 
of $A_v>7$ while CO becomes important at $A_v>4$, which shows that 
photodestruction could explain the offset between the CO and 
the H$_2$CO peaks. 

\section{Profile moments in detail}

\begin{deluxetable}{rrrrrc}
\tablewidth{0pt}
\tablecaption{Profile moments for each position of Fig. \ref{fig:mosaic}. The upper limits
correspond to $3 \sigma$. The symbol ``\ldots'' indicates no data were
available, while ``nf'' indicates that data were available but no
reliable fit could be made. Values marked with ``yes''
 in column 6 are
plotted as squares in Fig.\ \ref{fig:intensityratio}, and were
used for the least square fits. 
\label{table:mom}}
\tablehead{
\colhead{Offset (\textit{l,b})} &
\colhead{$1000 \times$ I(\htco)} & \colhead{$\langle V \rangle$} &
\colhead{I(CO)} & \colhead{$\langle V \rangle$} & \colhead{Included in} \\
\colhead{arcmin} &
\colhead{K \kmps} & \colhead{\kmps} &
\colhead{K \kmps} & \colhead{\kmps} & \colhead{fits}
}
\startdata
40, -70  & $<$ -24.3 & nf  & \ldots & \ldots   \\
30, -70  & $<$ -18.9 &  nf & \ldots     & \ldots \\
20, -70  & $<$ -25.8 & nf  & \ldots    & \ldots  \\
10, -70  & $<$ -23.4  & nf & 4.1 $\pm$ 0.7 & -7.6 $\pm$ 1.3  \\
0, -70   &-42.1 $\pm$ 8.4 & -7.5 $\pm$ 1.5 & 3.5 $\pm$ 0.6 &-8.7 $\pm$ 1.5 & yes \\
-10, -70 & -30.3 $\pm$ 7.1 & -6.2 $\pm$ 1.5 & 8.6 $\pm$ 0.8 & -8.3 $\pm$ 0.8 & yes \\
-20, -70 & $<$ -21.6  & nf  & $<$ 2.1  & nf &  \\
\tableline
40, -60  & $<$ -22.5 & nf & \ldots & \dots  \\
30, -60  & $<$ -24.0 & nf  &  \ldots & \ldots \\
20, -60  & $<$ -26.7 & nf & 5.9 $\pm$ 0.8 & -7.0 $\pm$ 1.1 \\
10, -60  & $<$ -23.7 & nf  & 5.1 $\pm$ 0.6 & -8.3 $\pm$ 1.0  \\
0, -60  &  $<$ -19.2 & nf & $<$ 1.8  & nf  \\
-10, -60 & -27.9$\pm$ 7.1 & -6.7 $\pm$ 0.5 & $<$ 2.1 & nf \\
-20, -60 & $<$ -28.8 & nf & $<$ 1.5 & nf  \\
\tableline
40, -50  & $<$ -25.5 &  nf & \ldots & \ldots  \\
30, -50 & -109.3 $\pm$ 8.9 &-2.6 $\pm$ 0.2 & 14.8 $\pm$ 1.1 & -1.8 $\pm$ 0.2 & yes \\
20, -50  & $<$ -25.2 & nf & 10.5 $\pm$ 0.8 & -5.0 $\pm$ 0.4  \\
10, -50  & $<$ -21.6 & nf & 8.2 $\pm$ 0.6 & -7.0 $\pm$ 0.5  \\
0, -50    & $<$ -18.9& nf & $<$ 2.4 & nf  \\
-10, -50 & $<$ -16.5   & nf & $<$ 2.4 & nf  \\
-20, -50 & $<$ -24.6  & nf & $<$ 4.8 & nf  \\
\tableline
40, -40    & $<$ -23.1 &nf & \ldots & \ldots \\
30, -40    & $<$ -22.8 &nf & 3.1 $\pm$ 0.6 & -6.2 $\pm$ 1.2   \\
20, -40    & $<$ -22.2  & nf & 7.2 $\pm$ 0.8 &-7.2 $\pm$ 0.9 \\
10, -40    & $<$ -23.1 & nf & 4.5 $\pm$ 0.5 & -8.8 $\pm$ 1.1  \\
0, -40     & $<$ -22.5    & nf & $<$ 2.7 & nf  \\
-10, -40   & $<$ -26.1 & nf & $<$ 2.7 & nf   \\
-20, -40   & $<$ -18.6 & nf & $<$ 3.0  & nf  \\
\tableline
40, -30    & $<$ -22.5 & nf & 3.4 $\pm$ 0.5 & -9.0 $\pm$ 1.5  \\
30, -30  & $<$ -22.8 & nf & 4.0 $\pm$ 0.7 & -8.9 $\pm$ 1.5 \\
20, -30    & $<$ -25.5 & nf & $<$ 1.8 & nf  \\
10, -30    & $<$ -30.9 & nf & 5.2 $\pm$ 0.6 &-8.6 $\pm$ 1.0  \\
0, -30     & $<$ -20.1 & nf & $<$ 4.2 & nf \\
-10, -30   & $<$ -26.7  & nf & $<$ 1.8 & nf  \\
-20, -30   & $<$ -22.2 & nf & $<$ 2.1 &nf  \\
\tableline
40, -20    & \ldots     & \ldots   &  $<$ 2.7 & nf  \\
30, -20  &  $<$ -30.9 &nf & 3.5 $\pm$ 0.7 &-9.4  $\pm$ 1.9  \\
20, -20    & $<$ -24.0 & nf & $<$ 2.7 & nf  \\
10, -20  & -31.6 $\pm$ 6.0 &-8.2 $\pm$ 1.6 & 8.2 $\pm$ 1.1 & -8.9 $\pm$ 1.2 & yes \\
0, -20     & $<$ -29.1  & nf & 8.5 $\pm$ 1.4 & -9.5 $\pm$ 1.7   \\
-10, -20   & $<$ -19.5  & nf & 6.8 $\pm$ 0.3 & -8.7 $\pm$ 0.9   \\
-20, -20   & $<$ -22.2 & nf & $<$ 1.5 & nf  \\
\tableline
40, -10     & \ldots   & \ldots & $<$ 1.8 & nf  \\    
30, -10     & $<$ -35.7 & nf &$<$ 2.4  &nf  \\   
20, -10 &  $<$ -16.8 & nf & $<$ 3.0  & nf  \\   
10, -10  &-43.1 $\pm$ 8.2 & -8.1 $\pm$ 1.6 & 4.7 $\pm$ 1.1 & -10.3 $\pm$ 2.8 & yes \\   
0, -10  & -76.2 $\pm$ 13.8 & -9.5 $\pm$ 1.8 & 11.7 $\pm$ 0.8 &-9.0 $\pm$ 0.6 & yes \\   
-10, -10  & -82.7 $\pm$ 7.7 & -8.1 $\pm$ 0.8 & 18.6 $\pm$ 0.6 & -8.2 $\pm$ 0.3 & yes \\   
-20, -10    & $<$ -24.3 & nf & $<$ 2.7 & nf \\   
\tableline
40, 0       & $<$ -33.0 & nf & 2.8 $\pm$ 0.5 & -11.6 $\pm$ 2.3   \\   
30, 0       & $<$ -28.8  &nf & $<$ 3.3 & nf  \\   
20, 0 & -63.8 $\pm$ 7.8 & -7.8 $\pm$ 0.9 & 3.8 $\pm$ 1.1 &-8.9 $\pm$ 3.0 & yes \\   
10, 0 &-93.7 $\pm$ 11.3 & -8.7 $\pm$ 1.1 & 9.4 $\pm$ 0.7 & -9.9 $\pm$ 0.8 & yes \\  
 0, 0 & -272.6 $\pm$ 9.8 &-7.8 $\pm$ 0.3 & 21.6 $\pm$ 0.7 & -8.0 $\pm$ 0.3 &  yes \\  
-10, 0 & -159.8 $\pm$ 8.6 &-8.0 $\pm$ 0.4 & 46.5 $\pm$ 1.6 & -7.8 $\pm$ 0.3 & yes  \\  
-20, 0      & $<$ -24.0     & nf & 10.2 $\pm$ 0.7 & -7.7 $\pm$ 0.6  \\   
\tableline
40,+10      & \ldots   & \ldots  & $<$ 1.7  & nf  \\   
30, +10     & $<$ -28.8 & nf & $<$ 0.9 & nf  \\
20, +10  & -52.0 $\pm$ 8.2 &-8.7 $\pm$ 1.4 & 5.3 $\pm$ 0.4 &-9.1 $\pm$ 0.7  & yes \\
10, +10 & -114.1 $\pm$ 9.7 &-6.2 $\pm$ 0.6 & 11.1 $\pm$ 1.3 & -7.1 $\pm$ 0.9 & yes  \\
                 & -70.3 $\pm$ 7.7  & -10.3 $\pm$ 1.2 & 13.5 $\pm$ 1.1 & -10.6 $\pm$ 1.0 & yes \\
0, +10 & -247.0 $\pm$ 10.1 & -7.6 $\pm$ 0.3 & 22.8 $\pm$ 1.6 &-8.5 $\pm$ 0.6 & yes \\
-10, +10 &-118.9 $\pm$ 9.2 & -8.1 $\pm$ 0.6 & 22.3 $\pm$ 2.0 &-8.6 $\pm$ 0.8 & yes \\
-20, +10    & $<$ -19.2 & nf & $<$ 0.9 & nf  \\
\tableline
40, +20      & \ldots   & \ldots  & $<$ 1.2 &nf  \\ 
30, +20      & $<$ -25.2 & nf & 3.1 $\pm$ 0.4 & -8.3 $\pm$ 1.7  \\
20, +20 & -42.3 $\pm$ 9.3 & -7.1 $\pm$ 1.6 & 6.7 $\pm$ 0.7 & -9.9 $\pm$ 1.0 &yes \\
                 &-82.0 $\pm$ 7.4   & -11.3 $\pm$ 1.1 & nf & nf  \\
10, +20 & -49.8 $\pm$ 6.9  &-6.9 $\pm$ 1.0 & 7.0 $\pm$ 1.2 & -8.0 $\pm$ 1.4 &yes \\
                 &-63.1 $\pm$ 5.5 & -10.8 $\pm$ 1.0 & 8.3 $\pm$ 0.9 &-11.0 $\pm$ 1.4 & yes \\
0, +20 &-68.3 $\pm$ 9.7 & -8.9 $\pm$ 1.3 & 6.1 $\pm$ 0.7 & -10.8 $\pm$ 1.3 &yes \\
-10, +20 &-56.4 $\pm$ 6.8 &-8.6 $\pm$ 1.1 & 5.5 $\pm$ 0.5 & -10.8 $\pm$ 1.1 & yes \\
-20, +20     & $<$ -29.7  & nf &  $<$ 2.7 & nf  \\
\tableline
40, +30      & \ldots   & \ldots  &  $<$ 1.6 & nf  \\
30, +30      & $<$ -27.9  &nf & $<$ 1.6  &nf \\
20, +30      & $<$ -19.5 & nf & 3.4 $\pm$ 0.8 & -11.4 $\pm$ 2.7  \\
10, +30  & -54.1 $\pm$ 8.8 & -9.3 $\pm$ 1.6 & 5.6 $\pm$ 0.7 & -10.3 $\pm$ 1.2 & yes \\
0, +30       & $<$ -18.0 & nf & 5.0 $\pm$ 0.6 & -11.1 $\pm$ 1.3  \\
-10, +30     &$<$ -27.0 &nf& 2.0 $\pm$ 0.4 & -10.2 $\pm$ 2.2  \\
-20, +30     & $<$ -23.4 & nf & $<$ 3.0 & nf  \\
\enddata
\end{deluxetable}

\end{document}